\documentclass[a4paper,11pt]{article}

\usepackage{jcappub}

\usepackage[T1]{fontenc}
\usepackage[utf8]{inputenc}
\usepackage{amsmath}
\usepackage[mathscr]{euscript}
\usepackage{graphicx}

\newcommand{\la}{\langle}
\newcommand{\ra}{\rangle}
\newcommand{\lp}{\left(}
\newcommand{\rp}{\right)}
\newcommand{\x}{\mathbf{x}}
\renewcommand{\k}{\mathbf{k}}
\newcommand{\kh}{\hat{k}}
\newcommand{\q}{\mathbf{q}}
\newcommand{\qp}{\mathbf{q'}}
\renewcommand{\d}{\delta}
\newcommand{\dk}{\d^K} 
\newcommand{\f}{\frac}
\newcommand{\dd}{\partial} 
\newcommand{\sym}{\mathrm{sym}} 

\newcommand{\rhobar}{\bar{\rho}_{m,0}}
\newcommand{\dm}{\d_m} 
\newcommand{\dg}{\d_g} 
\newcommand{\dlin}{\d^{\mathrm{lin}}}
\newcommand{\sh}{\gamma^I} 
\newcommand{\shij}{\gamma^I_{ij}} 
\newcommand{\wsh}{\tilde{\gamma}^I} 

\newcommand{\sxs}{s \otimes s}
\newcommand{\Plin}{P^{\mathrm{lin}}} 
\renewcommand{\xi}{\x_{\mathrm i}} 
\newcommand{\xf}{\x_{\mathrm f}} 
\newcommand{\ti}{t_{\mathrm i}} 
\newcommand{\tf}{t_{\mathrm f}} 
\newcommand{\zi}{z_{\mathrm i}} 
\newcommand{\zf}{z_{\mathrm f}} 
\newcommand{\zia}{z_{\mathrm{IA}}} 
\newcommand{\eff}{_{\mathrm{eff}}} 
\newcommand{\sgam}{\sigma_{\gamma}} 
\newcommand{\nbar}{\bar{n}} 

\newcommand{\fdq}{\f{\mathrm d^3 \mathbf q}{(2 \pi)^3}}
\newcommand{\conn}{_{\mathrm c}}
\newcommand{\tri}{\bigtriangleup}
\newcommand{\ggi}{\emph{ggI }}

\newcommand{\ld}{\mathscr D} 

\DeclareMathOperator{\cov}{Cov} 
\DeclareMathOperator{\conv}{conv} 

\definecolor{jabgreen}{RGB}{0,192,0}

\title{\boldmath Time evolution of intrinsic alignments of galaxies}

\author[a,1]{D. M. Schmitz,\note{Corresponding author.}}
\author[b]{C. M. Hirata,}
\author[b,c]{J. Blazek,}
\author[a,d]{E. Krause}

\affiliation[a]{TAPIR, Mailcode 350-17, California Institute of Technology, Pasadena, CA 91125, USA}
\affiliation[b]{Department of Physics, The Ohio State University\\191 West Woodruff Ave.,
Columbus, Ohio 43210, USA}
\affiliation[c]{Laboratory of Astrophysics, \'Ecole Polytechnique F\'ed\'erale de Lausanne, CH-1290 Versoix, Switzerland}
\affiliation[d]{Jet Propulsion Laboratory, California Institute of Technology, 4800 Oak Grove Dr., Pasadena, CA 91109, USA}

\emailAdd{ds@astro.caltech.edu}
\emailAdd{hirata.10@osu.edu}
\emailAdd{blazek@berkeley.edu}
\emailAdd{ekrause@caltech.edu}

\abstract{Intrinsic alignments (IA), correlations between the intrinsic shapes and orientations of galaxies on the sky, are both a significant systematic in weak lensing and a probe of the effect of large-scale structure on galactic structure and angular momentum. In the era of precision cosmology, it is thus especially important to model IA with high accuracy. Efforts to use cosmological perturbation theory to model the dependence of IA on the large-scale structure have thus far been relatively successful; however, extant models do not consistently account for time evolution. In particular, advection of galaxies due to peculiar velocities alters the impact of IA, because galaxy positions when observed are generally different from their positions at the epoch when IA is believed to be set. In this work, we evolve the galaxy IA from the time of galaxy formation to the time at which they are observed, including the effects of this advection, and show how this process naturally leads to a dependence of IA on the velocity shear. We calculate the galaxy-galaxy-IA bispectrum to tree level (in the linear matter density) in terms of the evolved IA coefficients. We then discuss the implications for weak lensing systematics as well as for studies of galaxy formation and evolution. We find that considering advection introduces nonlocality into the bispectrum, and that the degree of nonlocality represents the memory of a galaxy's path from the time of its formation to the time of observation. We discuss how this result can be used to constrain the redshift at which IA is determined and provide Fisher estimation for the relevant measurements using the example of SDSS-BOSS.}


\begin{document}
\maketitle
\flushbottom

\section{Introduction}
\label{sec:intro}

Weak lensing (WL) is a powerful probe of large-scale structure (LSS) and dark energy \cite{blandford1991, miralda-escude1991, kaiser1992, refregier2003, kilbinger2015}, especially in combination with other cosmological measurements \cite{yoo&seljak2012}. Surveys designed to measure WL, such as COSMOS, CFHTLenS, DES, and KiDS, have already achieved success in constraining cosmological parameters \cite{descollaboration2017, hoekstra2006, huff2014, jee2013, jee2016, joudaki2018, kilbinger2013, massey2007, mandelbaum2013, miyatake2015, schrabback2010, semboloni2011} and testing general relativity and modified gravity models \cite{daniel2010, lombriser2012, reyes2010, tereno2011, thomas2009}, while future WL surveys will improve upon these constraints \cite{laureijs2011, lsstcollab2012, spergel2013} and perhaps even provide independent constraints on CMB B-mode polarization due to primordial gravitational waves \cite{chisari2014, schmidt&jeong2012}.

WL requires large samples of galaxies in order to reduce the relative contribution from random ``shape noise'' -- the intrinsic (\emph{i.e.}, unlensed) ellipticities of galaxies are much larger than the lensing shear. If the shape noise is independent for each galaxy, then it does not bias the correlation function at nonzero lag. However, the intrinsic galaxy shape field is not purely random, but includes a correlated component known as the intrinsic alignments (IA) \cite{crittenden2001, hirata&seljak2004, joachimi2015, troxel&ishak2015} described by the intrinsic alignment tensor $\sh_{ij}(\x)$. It is therefore extremely important \cite{kirk2012, krause2016} to model IA accurately in order to mitigate the associated systematic errors in WL \cite{laszlo2012} and redshift-space distortion \cite{hirata2009, martens2018} studies. Because IA are impacted by the large-scale tidal field and other fields, they are also potentially of interest as a probe of the LSS and its effects on galaxies. This effect has already been studied in simulations \cite{chisari2015, hilbert2017, tenneti2015} and observationally, on small scales in highly clustered environments \cite{godlowski2012, rong2015}, for red galaxies in SDSS-BOSS \cite{blazek2012, okumura&jing2009, singh2015} and Mega-Z \cite{joachimi2011}, and for blue galaxies in CFHTLenS \cite{tonegawa2017} and WiggleZ \cite{mandelbaum2011}. Observations suggest as well that different galaxy populations exhibit different magnitudes of the IA signal, with more massive galaxies aligning more strongly with the LSS \cite{li2013} and different color/morphological types displaying different IA amplitudes, underscoring the need for robust models of the physical processes which influence IA.

Previous work has examined two primary models for the effect of the tidal field on IA. The \emph{linear alignment} or \emph{tidal alignment} model \cite{catelan2001} posits that the axes of a triaxial galaxy are preferentially aligned with the axes of the tidal field, and in particular, that the long axis of the ellipsoid is preferentially aligned parallel to the stretching axis of the tidal quadrupole. The \emph{quadratic alignment} or \emph{tidal torque} model \cite{mackey2002, hui&zhang2002, schaefer&merkel2012} represents a second-order contribution due to tidal torquing. In this model, the formation of an angular momentum axis accounts for one linear power of the tidal field, and the resulting torque on this axis accounts for the second. The linear model is the dominant effect on IA in large elliptical galaxies, and correctly predicts the scale dependence in the linear regime \cite{hirata2007}, whereas the quadratic model is believed to be more relevant to disk galaxies \cite{catelan2001}. Note, however, that due to loop corrections in non-linear perturbation theory, a quadratic model valid on small scales results in the appearance of a linear term on large scales \cite{hui&zhang2002, blazek2017}. More generally, these contributions can be considered part of an effective expansion in all potentially relevant cosmological fields at a given order \cite{blazek2017}.

In this work, we generalize this expansion to consider all linearly independent quantities that contribute at second order in the linear matter overdensity $\dlin$, including both the linear and quadratic model contributions. Following the approach to galaxy clustering taken in \cite{mcdonald&roy2009}, we decompose the intrinsic alignment field into components depending on these quantities in a manner analogous to the use of bias coefficients to quantify clustering. To second order, there are four such terms, which we shall describe in Section~\ref{sec:calc}. The full second-order standard perturbation theory (SPT) model can then be used to predict the galaxy density-galaxy density-intrinsic alignment ($ggI$) bispectrum to tree level (fourth order in the linear matter density). Although bispectra are not as widely used as power spectra, they are of great interest since they can improve LSS constraints in combination with 2-point statistics (\emph{e.g.}, \cite{sefusatti2006}). The variety of configuration dependences of the bispectrum can be used to break degeneracies present in the power spectrum alone. Moreover, the tree-level bispectrum is a straightforward early step in the theory of non-linear biasing: it can be computed using biasing terms through order $(\dlin){^2}$, and such terms contribute at leading order. The non-linear corrections to the power spectrum would require terms through order $(\dlin){^3}$; we shall investigate these in a future paper.

It is frequently assumed that the bulk of the effect of the tidal field on IA occurs at high redshift (and is associated with the galaxy formation process) \cite{camelio&lombardi2015}, around the time of structure formation and initial baryonic collapse. This assumption has thus far been difficult to probe observationally due to degeneracies with IA amplitude (\emph{e.g.}, discussion in \cite{blazek2015}). Lensed galaxies in a WL measurement are typically observed at an intermediate redshift ($z \lesssim 1.5$), and measurements of IA relate galaxy alignments to the cosmological fields at the observed redshift. This fact complicates IA modeling because it means that observers today measure different bias coefficients from those that would be measured by a hypothetical observer at the formation redshift; in fact, the time evolution of IA has been detected in hydrodynamic simulations \cite{chisari2016}. There are two reasons for the discrepancy, even if the IA evolution is passive (\emph{i.e.}, a galaxy simply maintains its shape after the formation redshift). First, the cosmological fields evolve over time, so the bias coefficients must be normalized to a different value at each time point to take this evolution into account. Second, the peculiar velocity of each galaxy results in advection, \emph{i.e.}, the position of a galaxy at a low redshift $\zf$ is different from the position of the same galaxy at a higher redshift $\zi$. In this work, we include these time-evolution effects in order to better characterize the IA contamination to WL observables, as well as to outline how IA observations may be used to constrain galaxy formation itself.

This work is organized as follows. In Section~\ref{sec:calc}, we develop a formalism for decomposing IA in terms of bias coefficients and analytically determine how these bias coefficients evolve in time. In Section~\ref{sec:num}, we present the \ggi bispectrum for a set of fiducial values of the bias coefficients, and describe how the time evolution of the coefficients can be used to probe galaxy formation physics. In Section~\ref{sec:error}, we perform Fisher information matrix analysis to determine how well the coefficients can be measured in SDSS-BOSS data and how well the redshift of galaxy IA determination can be constrained based on the time evolution of the bias coefficients. In Section~\ref{sec:disc}, we discuss these results and future research directions.

Throughout this work, we work in comoving Mpc$/h$ units and adopt the cosmology of \emph{Planck} (2015) \cite{planckcollaboration2016}.

\section{Calculations}
\label{sec:calc}

The objective of IA modeling on linear and quasi-linear scales is to express the IA field $\sh(\x)$ as a function of cosmological fields and IA ``bias'' coefficients. In this work, we compute $\sh$ to second order in $\dlin$ so that the \ggi bispectrum can be obtained to fourth order (tree level). Galaxy formation models concern the galaxy properties at the position and redshift of formation, whereas observations measure these properties at the position and redshift of observation. This fact motivates the development in this work of an expression for the time evolution of $\sh$ from initial coordinates $(\xi,\zi)$ to final coordinates $(\xf, \zf)$. 

In this section, we shall consider the effects of the advection $\xi \rightarrow \xf$ as well as the time evolution of the cosmological fields, incorporate the time evolution into a calculation of $\sh$ in Eulerian SPT, and thus derive a system of equations for the passive evolution of an IA field. Finally, we shall calculate the \ggi bispectrum taking these effects into account.

In what follows, $\d_m(\x,t)$ denotes the matter density perturbation at Eulerian position $\x$ and time $t$. The matter density $\d_m$ is expanded to second order in $\dlin$ in the usual way:
\begin{equation}\d_m (\x,t) \approx \dlin(\x) D(t) + \d^{(2)} (\x) D(t)^2 + \ldots,
\end{equation}
where $D(t)$ is the linear growth factor, and $\d^{(2)}$ is the second-order density perturbation defined by a convolution with the $F_2$ kernel (\emph{e.g.}, \cite{goroff1986, mukhanov1992, bernardeau2002}): \begin{equation} F_2(\q_1, \q_2) = \f 57 + \f 12 \f{\q_1 \cdot \q_2}{q_1 q_2} \lp \f{q_1}{q_2} + \f{q_2}{q_1} \rp + \f 27 \f{(\q_1 \cdot \q_2)^2}{q_1^2 q_2^2}.\end{equation}
We neglect the time dependence of $F_2$ because the kernel has no time dependence to second order in the Einstein-de Sitter (EdS) cosmology, and the deviation from perfect time-invariance in our adopted cosmology is sufficiently small that it can be neglected without consequence.

\subsection{Formalism and bias coefficients for IA}
\label{subsec:formalism}

The intrinsic shear tensor describing the unlensed ellipticity of a galaxy in the $x - y$ plane is given by \begin{equation} \sh = \f{1}{2R} \lp \f{1 - (b/a)^2}{1 + (b/a)^2} \rp \begin{pmatrix} \cos 2 \theta & \sin 2 \theta \\ \sin 2 \theta & -\cos 2 \theta \end{pmatrix} \label{eq:defgamma} \end{equation} where $R$ is the shear responsivity (see \cite{bernstein&jarvis2002}), $b/a$ is the galaxy axis ratio, and $\theta$ is the position angle of the major axis measured with respect to the $x$-axis. As discussed below, this expression is a projection on the sky of the more generic traceless-symmetric tensor that describes galaxy shapes in 3D (\emph{e.g.}, \cite{schmidt2015}). The observed intrinsic shear of an individual unlensed galaxy includes both a stochastic component (shape noise) and a correlated component (intrinsic alignments). The shape noise contribution has no impact on correlation functions containing a single IA field (\emph{e.g.}, $gI$ or $ggI$) since it averages to zero, and because it is uncorrelated it has no effect on {\em any} correlation functions at non-zero lag. In the work that follows, we therefore neglect shape noise and use the notation $\sh$ to refer exclusively to the \emph{correlated} (\emph{i.e.}, determined by LSS) component of the intrinsic shear.

In general, we would like to express the IA field in terms of a series expansion in the linear matter density $\dlin$, in a manner analogous to the expression for galaxy biasing given by \emph{e.g.}, \cite{mcdonald&roy2009}:
\begin{equation}
\d_g (\x,t) \approx b_1(t) \d_m(\x,t) + \f{1}{2} b_2(t) \d_m^2(\x,t) + \f{1}{2} b_{s^2}(t) s_{ij} s_{ij} (\x,t) + \ldots\,. 
\label{eq:bias}
\end{equation}
That is, we shall write the expansion in terms of some number of ``bias'' coefficients multiplying combinations of the cosmological fields. To compute the \ggi bispectrum to tree level, we require $\sh$ to second order in $\dlin$. The relevant question is then in what ways $\dlin$ can enter the expression for $\sh$ up to second order; for example, it is known that galaxy biasing depends on the tidal field in addition to the density field alone \cite{baldauf2012}. We approach this question following the example of \cite{mcdonald&roy2009}. The approach taken in that work can be roughly summarized as follows. If galaxy formation were independent of the history of the density and velocity fields, then $\d_g(\x, z)$ should only depend on the matter overdensity $\dm(\x, z)$, which sources the gravitational potential $\Phi$ via the Poisson equation, and the peculiar velocity divergence $\theta = \nabla_i v_i(\x, z)$.\footnote{In this work, we assume all matter species behave as a single fluid, neglecting terms such as relative velocity between dark matter and baryons (\emph{e.g.}, \protect{\cite{tseliakhovich&hirata2010, blazek2016}}).} However, allowance for time evolution necessarily implies that additional terms appear in Eq.~\ref{eq:bias} \cite{chan2012, matsubara2011, saito2014}; these additional terms can be non-local in the sense that they do not depend on the instantaneous density, velocity gradient, and tidal field at $\x$. All such terms are restricted by the gauge invariance and the equivalence principle: $\d_g$ cannot depend on the zero-point of $\Phi$, nor on the overall addition of a constant gravitational field $\nabla_i\Phi$. Also, rotational invariance for the galaxy over-density implies that only scalar quantities can appear on the right-hand side of Eq.~\ref{eq:bias}.

Similar considerations apply when writing a biasing expansion of the intrinsic shear $\sh$. In this case, only terms that have the symmetry of a traceless-symmetric tensor (\emph{i.e.}, the same as $\sh$ itself) will contribute. This is analagous to imposing the condition that the bias coefficients for galaxy density are rotationally invariant, (\emph{i.e.}, are scalars). By inspection, we can find four independent terms with quadrupole symmetry that contribute at second order, as follows:
\begin{align}
\label{eq:tidal} s_{ij}(\x, z) &= - (1 + z) \lp \nabla_i \nabla_j \nabla^{-2} - \f{1}{3} \dk_{ij} \rp \d(\x, z), \\
\label{eq:ttorque} (\sxs)_{ij}(\x, z) &= s_{ik} (\x, z) s_{jk}(\x, z) - \f{1}{3} \dk_{ij} s_{kl}(\x, z) s_{kl}(\x, z), \\
\label{eq:delta s} \d s_{ij}(\x, z) &= \d(\x, z) s_{ij}(\x, z),~~{\rm and} \\
\label{eq:t} t_{ij}(\x, z) &= - (1 + z) \lp \nabla_i \nabla_j \nabla^{-2} - \f{1}{3} \dk_{ij} \rp \lp \nabla_k v_k (\x, z) - \d (\x, z) \rp.
\end{align}
(We argue in Appendix \ref{apx:lag} that these are the only four fields with the correct symmetry and order constructed from the density and tidal fields, and with no higher derivatives.)
Note that our normalization convention differs from that of \cite{mcdonald&roy2009} by the comoving factor $(1+z)$, and that we have defined $v$ as the peculiar velocity in comoving units. We have also normalized $t$ and $s$ in the preceding equations to eliminate the overall factor of $4 \pi G \rhobar$. That is, the physically relevant tidal field is defined as the traceless part of $\nabla_i \nabla_j \Phi$ where the (comoving) gravitational potential $\Phi$ is given by
\begin{equation}
\Phi(\x, t) = 4 \pi G \rhobar (1 + z(t)) \nabla^{-2} \d(x,t),
\end{equation} but it is more convenient to absorb these constants into the ``bias'' coefficients and explicitly include only the $z$-dependence. Note that this choice of $z$-dependence corresponds to the ``instantaneous alignment'' scenario \cite{blazek2015} in that the IA bias coefficients reflect dependences on the cosmological fields at the observed redshift. As discussed below, we are not assuming that this scenario is correct -- the amplitude and evolution of these coefficients will capture the true underlying IA history. The sign convention for $s$ has been chosen such that positive $s_{ij}\hat n_i\hat n_j$ corresponds to stretching along the $\hat{\mathbf n}$ direction (\emph{i.e.}, along the stretching axis), and the sign convention for $t$ has been chosen for consistency with $s$.

We have  defined $t$ such that, given the above normalizations, $t$ vanishes at linear order and is not degenerate with $\delta$ at second order. This choice of definition is physically relevant in addition to being convenient. The $t$ term defined in this manner encodes the degree of non-locality in $\sh$, in the sense that it is not determined by the instantaneous gravity gradient ($\nabla_i\nabla_j\Phi$; or alternatively $\d$ and $s_{ij}$) at point $\x$. What we describe as ``non-locality'' here could also be thought of as ``path-dependence,'' in the sense that the full history of the density and tidal field experienced by the galaxy are important; in perturbation theory, this information is still encoded in the full density field, just not all at point $\x$. The dependence on $\nabla_i v_j$ means that $t_{ij}$ encodes the memory of a galaxy's tidal history. We shall make this notion more precise in Section~\ref{subsec:adv}.

The intrinsic alignment field $\sh$ can then be expressed to second order in $\dlin$ as
\begin{equation}
\gamma^I_{ij}(\x, z) \approx c_s(z) s_{ij}(\x, z) + c_{\sxs}(z)(\sxs)_{ij}(\x, z) + c_{\d s}(z) \d s_{ij}(\x, z) + c_t(z) t_{ij}(\x, z).
\end{equation}  
This form suggests that we think of $\sh$ as a ``vector'' in a space spanned by the basis vectors $s_{ij}$, $\sxs_{ij}$, $\d s_{ij}$, and $t_{ij}$ (see \cite{schmidt2015,blazek2017}, where similar expansions were introduced without in including contributions from $t_{ij}$). The $c$ coefficients represent the components of $\sh$ in this space. The question addressed in this work -- how the observable IA at low redshift are related to galaxy formation physics at high redshift -- is therefore equivalent to the question of how vectors in this vector space transform under the time evolution operator.

\subsection{Passive evolution and the time evolution matrix}
\label{subsec:matrix}

We define {\em passive evolution} to be the scenario where the IA of an individual galaxy is determined at some point during galaxy formation and subsequently remains fixed. That is, the intrinsic shear of a galaxy at a later time ($z=\zf$) is equal to that of the same galaxy at an earlier time ($z=\zi$): $\gamma^I_{\rm L}({\mathbf q},\zf) = \gamma^I_{\rm L}({\mathbf q},\zi)$, where ${\mathbf q}$ denotes a Lagrangian position and the subscript ``L'' denotes a Lagrangian function. Of course, in the Eulerian description the $\gamma^I$ field will evolve due to advection.

In the case of passive evolution, the bias coefficients at $\zf$ can be expressed in terms of their values at $\zi$. We may thus write a time evolution operator $T(\zi \rightarrow \zf)$ in matrix form. Considering terms through second order in $\dlin$, we shall have
  	\begin{equation} 
  	\begin{pmatrix}
  	c_s(\zf) \\
  	c_{\sxs}(\zf) \\
  	c_{\d s}(\zf) \\
  	c_t(\zf) \end{pmatrix} = \begin{pmatrix}
  	T_{s,s} & 0 & 0 & 0 \\
  	T_{\sxs,s} & T_{\sxs, \sxs} & 0 & 0 \\
  	T_{\d s,s} & 0 & T_{\d s,\d s} & 0 \\
  	T_{t,s} & 0 & 0 & T_{t,t} \end{pmatrix} \begin{pmatrix}
  	c_s(\zi) \\
  	c_{\sxs}(\zi) \\
  	c_{\d s}(\zi) \\
  	c_t(\zi) \end{pmatrix},
    \label{eq:T-matrix}
  	\end{equation} 
where we have written only the components of the ${\bf T}$ matrix that will turn out to be non-zero. Note that this expression is fully general in $\zi$ and $\zf$: in particular, $\zi$ can be any redshift higher than $\zf$ and need not have any special significance to galaxy formation.

The diagonal elements of the matrix $T(\zi \rightarrow \zf)$ are simple time-evolution factors accounting for the fact that the density and tidal fields change over time, so the observed dependence of $\sh$ on the fields will change if $\sh$ remains fixed. The elements in the leftmost column are the result of nonlinear effects such as advection. The $s$ term is the only term that appears in the expression for $\sh$ to first order; the other bias terms are intrinsically second-order. Because the advection involves a power of $\dlin$, the $s$ term is the only term for which advection will contribute to $\sh$ at second order. Therefore, off-diagonal terms appear only in the $c_s$ column at this order. We shall find that the advection mixes up the terms, introducing new contributions from $c_s$ into the other terms. 

A variant on the idea of passive evolution is that the internal dynamics of galaxies might slowly randomize their orientations as a function of time, thereby reducing correlations with large scale structure. If this process is internal to the galaxy and independent of large-scale structure, this would be described by
\begin{equation}
\gamma^I_{\rm L}({\mathbf q},\zf) = A(\zi,\zf)\gamma^I_{\rm L}({\mathbf q},\zi)
+ {\rm [uncorrelated~noise]};
\end{equation}
the factor $A(\zi,\zf)\le 1$ describes how much memory of the original orientation remains at $z=\zf$. This idea would result in the multiplication of the ${\bf T}$ matrix in Eq.~\ref{eq:T-matrix} by a trivial factor of $A(\zi,\zf)$.

\subsection{Advection and time evolution}
\label{subsec:adv}

We are now ready to apply cosmological perturbation theory to determine the time evolution matrix.
The matrix elements in the leftmost column of the transformation matrix can be worked out by considering the two processes that occur in passive evolution: \emph{(i)} advection, and \emph{(ii)} the cosmological evolution of the density fields. We may write the Eulerian passive evolution equation as
\begin{equation} \gamma^I(\xf,\zf) = \gamma^I(\xi,\zi) = \gamma^I(\xf,\zi) + \gamma^I|_{\rm adv}(\xf,\zf),\label{eq:pe}
\end{equation}
where ${\bf x}_i$ is the Eulerian position of a particle at redshift $\zi$ that will advect to ${\bf x}_f$ at redshift $\zf$, and we have denoted the advection correction by $\gamma^I|_{\rm adv}$. This can be written as a Taylor expansion in the displacement:
\begin{equation} \sh_{ij}|_{\rm adv}(\xf,\zf) = \nabla_k \sh_{ij}(\xf,\zi) \cdot (\xi-\xf)_k + \ldots \,.
\label{eq:gij} \end{equation}
Since we want an expression for $\gamma^I$ to second order in $\dlin$, the only contribution to Eq.~\ref{eq:gij} will be via first-order advection in the first-order $s$ term. The first-order advection can be computed from the Zel'dovich approximation \cite{zeldovich1970, sahni1995}:
\begin{equation} (\xi - \xf)_k = \int_{t(\zf)}^{t(\zi)} v_k (\xf,t) \, dt = (D(\zf) - D(\zi)) \nabla_k \nabla^{-2} \dm(\xf), \end{equation} where $D(z)$ is the linear growth function, $t(z)$ is the age of the Universe at redshift $z$, and all equalities are valid to linear order. We then conclude that to second order,
\begin{equation} \gamma^I_{ij}|_{\rm adv}(\x,\zf) = c_s(\zi) \nabla_k s_{ij}(\x,\zi) (D(\zf) - D(\zi)) \nabla_k \nabla^{-2} \dm(\x).
\end{equation} After transforming to Fourier space, and defining $\q' \equiv \k - \q$, we find that the advection contribution is equal to \begin{align} \gamma^I_{ij}|_{\rm adv}(\k,\zf) = &-(1+\zi) c_s(\zi) D(\zi) (D(\zf) - D(\zi)) \nonumber \\ &\times \int  \lp \f{q_i q_j}{q^2} - \f{1}{3} \dk_{ij} \rp \f{\q \cdot \q'}{\q'^2} \dlin(q) \dlin(q') \fdq. \label{eq:adv} \end{align}
This term can then be decomposed into the four terms previously discussed above: $s_{ij}(\zf)$, $(\sxs)_{ij}(\zf)$, $\d s_{ij}(\zf)$, and $t_{ij}(\zf)$, which together with Eq.~\ref{eq:pe} give a complete description of $\shij$ to second order in $\dlin$. The details of this calculation are provided in Appendix~\ref{apx:decom}. This decomposition, along with the generic time dependence of the cosmological fields, determines the matrix elements of $T(\zi \rightarrow \zf)$. The nonzero matrix elements are as follows:
\begin{align}
T_{s,s} &= \f{(1+\zi)D(\zi)}{(1+\zf)D(\zf)}, \label{eq:matr-first}\\
T_{\sxs, \sxs} &= \f{(1+\zi)^2D(\zi)^2}{(1+\zf)^2D(\zf)^2},\\
T_{\d s,\d s} &= \f{(1+\zi)D(\zi)^2}{(1+\zf)D(\zf)^2},\\
T_{t,t} &= \f{(1+\zi)D(\zi)^2}{(1+\zf)D(\zf)^2},\label{eq:matr-last-diag} \\
T_{\sxs, s} &= \f{(D(\zf) - D(\zi))(1+\zi)D(\zi)}{(1+\zf)^2D(\zf)^2},\label{eq:matr-first-odiag}\\
T_{\d s, s} &= - \f 23 \f{(D(\zf) - D(\zi))(1+\zi)D(\zi)}{(1+\zf)D(\zf)^2},~~{\rm and}\\
T_{t,s} &= \f 52 \f{\lp D(\zf) - D(\zi) \rp (1 + \zi) D(\zi)}{(1 + \zf) D(\zf)^2}.\label{eq:matr-last}
\end{align}
Of particular interest is $T_{t,s}$, which denotes the contribution to the IA from $t_{ij}$. Even if galaxy formation is assumed to be local, setting $c_t(\zi) = 0$, advection introduces nonlocality into IA measurements such that $c_t(\zf) \neq 0$. This is discussed in detail in Section~\ref{subsec:nonlocal}.

An interesting feature of the time evolution matrix is that the $T_{s,s}$ component is first-order; the numerator and denominator both contain one growth factor $D$. The coefficient $c_s$ transforms generically in this way even though it applies to both the first-order and second-order components of the tidal field.

\subsection{\ggi bispectra} 
\label{subsec:bispec}

We now compute the \ggi bispectrum in terms of the IA bias coefficients evaluated at the final (observed) redshift. The bispectrum calculated here is defined by the Fourier transform of the galaxy density-galaxy density-galaxy weighted IA three-point correlation function:
\begin{eqnarray}
(2 \pi)^3 \d_D^3(\k_1 + \k_2 + \k_3) B^{ggI}(\k_1, \k_2, \k_3) &=& 
\int \la \dg(\x_1) \dg(\x_2) \wsh(\x_3)\ra
e^{-i\k_1\cdot\x_1}
e^{-i\k_2\cdot\x_2}
\nonumber \\ && \times
e^{-i\k_3\cdot\x_3}
\,d^3\x_1\,d^3\x_2\,d^3\x_3,
\end{eqnarray}
where the galaxy density weighted IA is the observed quantity defined by
\begin{equation} \wsh(\x) = \lp 1 + \d_g(\x) \rp \sh(\x). \end{equation}
 
\noindent In this work, we restrict our analysis to triangles in the plane of the sky (\emph{i.e.}, $\k_i$ perpendicular to the line of sight), since this is the case relevant to contamination of WL measurements. In this case, of the five components of the traceless-symmetric tensor $\sh_{ij}(\k)$, the observed ellipticities correspond to two of them. If we choose the coordinate system so that the $z$-direction is along the line of sight and the $x$-direction is along $\k$, we may write
\begin{equation}
\wsh_E(\k) = \frac{\wsh_{xx}(\k) - \wsh_{yy}(\k)}{2}
~~~{\rm and}~~~
\wsh_B(\k) = \wsh_{xy}(\k).
\end{equation}
We shall only consider these two components in what follows.\footnote{The $\wsh_{zz}$ component would be relevant to radial intrinsic alignments \protect{\cite{martens2018}}.}

These bispectrum modes can be determined in terms of the $c$ coefficients at redshift $\zf$, which are related to the values at the initial (formation) redshift via $T(\zi \rightarrow \zf)$. To simplify notation, let us define $D^4 \equiv D(\zf)^4$, $\mu_{nm} \equiv \k_n \cdot \k_m/k_n k_m$, and $\kh_{mi} = \k_m \cdot \hat{\mathbf e}_i/k_m$. We further follow \cite{hirata&seljak2004} and define the E-mode and B-mode kernels by
\begin{align} f_E(\k) &= \frac 12 \lp \kh_x^2 - \kh_y^2 \rp, \\
f_B(\k) &= \kh_x \kh_y, \\
h_E^{\sym} (\k_m, \k_n) &= \f{\mu_{mn}}{2} \lp \kh_{mx} \kh_{nx} - \kh_{my} \kh_{ny} \rp  - \f 13 \lp f_E(\k_m) + f_E(\k_n) \rp,~~{\rm and}\\
h_B^{\sym} (\k_m, \k_n) &= \f{\mu_{mn}}{2} \lp \kh_{mx} \kh_{ny} + \kh_{my}\kh_{nx} \rp - \f 13 \lp f_B(\k_m) + f_B(\k_n) \rp.
\end{align}
In the case where analysis is restricted to triangles in the plane of the sky -- that is, $\k_1$ and $\k_2$ are in the plane of the sky, and not just $\k$ -- we have:

\begin{align} h^{\sym}_E(\k_1, \k_2) &= \f 16 (f_E(\k_1) + f_E(\k_2)) \label{eq:hE}~~~{\rm and}
\\ h^{\sym}_B(\k_1, \k_2) &= \f 16 (f_B(\k_1) + f_B(\k_2))\label{eq:hB}.
\end{align}
(This simplification is not available for general $\k_1$ and $\k_2$.)
The two-argument kernels $h_E$ and $h_B$ appear in the expressions for the $\sxs$ term of the E-mode and B-mode \ggi bispectra.

We now write out the $E$-mode and $B$-mode bispectra. For simplicity, we suppress the time dependence of the IA and galaxy bias coefficients. All time-dependent quantities, including the growth function $D$, should be understood to be evaluated at $\zf$, except the linear power spectrum $P(\k)$, which should be understood to be evaluated at $z=0$. That is, the time dependence of the linear power spectrum has been made explicit in the form of prefactors of $D$.
\begin{align} B^{ggI}_E = &- c_s b_1 D^4 (1+z) \lp b_1 F_2^{\sym}(\k_2, \k_3) + b_2 + b_{s^2} (1+z)^2 \lp \mu_{23}^2 - \f 13 \rp \rp P(k_2)P(k_3) \nonumber\\
&-c_s b_1 D^4 (1+z) \lp b_1 F_2^{\sym}(\k_1, \k_3) + b_2 + b_{s^2} (1+z)^2 \lp \mu_{13}^2 - \f 13 \rp \rp P(k_1)P(k_3) \nonumber\\
&- c_s b_1^2 D^4 (1+z) \lp b_1 + 2 F_2^{\sym}(\k_1, \k_2) \rp P(k_1)P(k_2) \nonumber\\
&- \lp c_{\d s} - \f 13 (1+z) c_{\sxs} \rp b_1^2 D^4 (1+z) (f_E(\k_1) + f_E(\k_2)) P(k_1)P(k_2) \nonumber\\
&-\f 47 c_t b_1^2 D^4 (1+z) \lp \mu_{12}^2 - 1 \rp P(k_1)P(k_2) \label{eq:ebispec}~~{\rm and}\\
B^{ggI}_B = \, &- \lp c_{\d s} - \f 13 (1+z) c_{\sxs} \rp b_1^2 D^4 (1+z) (f_B(\k_1) + f_B(\k_2)) P(k_1)P(k_2). \label{eq:bbispec}
\end{align} 

\noindent Note that parity invariance does not require, in general, that the B-mode bispectrum vanishes; it requires only that $B^{ggI}_B(\k_1, \k_2, \k_3) = - B^{ggI}_B(R_y\k_1, R_y\k_2, R_y\k_3)$, where $R_y$ is a reflection operator.\footnote{This operator is defined by $R_y(k_x,k_y) = (k_x,-k_y)$.} That is, it requires that the bispectrum for the reflected triangle have a minus sign. This implies that the B-mode bispectrum vanishes for isosceles triangles with $k_1 = k_2$, because for such triangles the parity transformation is equivalent to a rotation, but this bispectrum can be nonvanishing for general triangles.

It is apparent from these equations that in an analysis restricted to triangles in the plane of the sky, the coefficients $c_{\sxs}$ and $c_{\d s}$ cannot be individually constrained; only the quantity $c_{\d s} - (1+z) c_{\sxs}/3$ can be constrained. This can be viewed as a consequence of Eqs.~\ref{eq:hE}~--~\ref{eq:hB}, which relate the angular kernels for the $\sxs$ terms to those for the $\d s$ terms. The existence of a degeneracy can also be deduced from group theory and symmetry arguments, which we outline in Appendix~\ref{apx:group}. 

\section{Numerical evaluation}
\label{sec:num}

In this section, we adopt the \emph{Planck} (2015) cosmology of \cite{planckcollaboration2016} as implemented in Astropy \cite{astropycollaboration2013} and use a linear matter power spectrum $\Plin(z=0)$ computed with CAMB \cite{lewis2000}. We also adopt survey parameters from SDSS-BOSS DR12 \cite{reid2016}. The survey consists of two subsamples, CMASS ($z\eff = 0.57$), and LOWZ ($z\eff = 0.32$). We combine the samples into a single BOSS sample with effective redshift taken to be a galaxy number-weighted average over the redshift distributions of both samples. This procedure gives $z\eff = 0.49$. 

We also require a measurement of $c_s$ in an LRG sample (any LRG sample should be suitable, because it represents a similar population of galaxies to those which we consider in this work) in order to choose fiducial values of the IA coefficients. We set $c_s(z = 0.32) = 0.1585 \, \Omega_m / ((1 + 0.32) D(0.32)) = 0.1421 \, \Omega_m$ in accordance\footnote{In this work, we normalize the growth function $D$ such that it is equal to 1 at $z=0$. The authors of \protect{\cite{blazek2011}} use the convention that $D = a$ during matter domination. We have converted their measurement to our conventions using $z=10$ as the normalization point for matter domination.} with observations of IA in the SDSS DR7 LRG sample \cite{blazek2011}. We have chosen this measurement because it was performed using isophotal shapes in order to maximize the observed IA signal. We normalize the time-varying galaxy bias value at $b_1(z = 0.49) = 1.8$ in accordance with SDSS-BOSS \cite{reid2016}, with $b_2 = 0.9(b_1 - 1)^2 - 0.5$ and $b_{s^2} = (4/7)(b_1 - 1)$.

\subsection{Bispectra}
\label{subsec:num_bispec}

\begin{figure}
\begin{center}
\includegraphics[width=\textwidth]{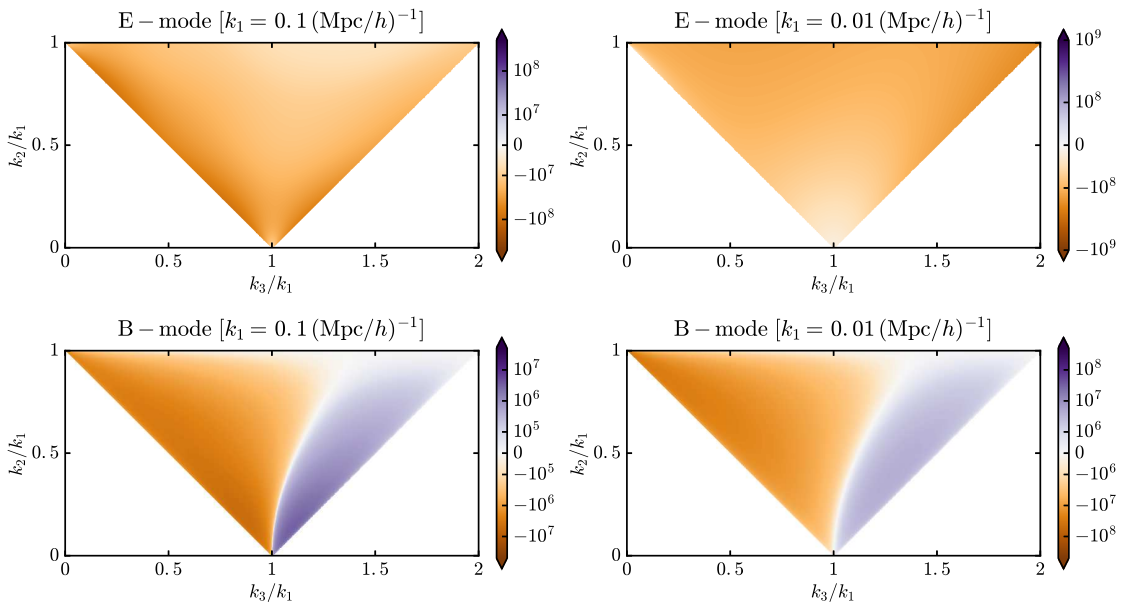}
\caption{The $E$-mode and $B$-mode \ggi bispectra in units of $(\mathrm{Mpc}/h)^6$ in configuration space for $k_1 = 0.1 \, \mathrm{Mpc}/h$ and $k_1 = 0.01 \, \mathrm{Mpc}/h$. These bispectra have been computed assuming the BOSS combined effective redshift $\zf = 0.49$ and a fiducial value of $\zi = 6$, with $c_s$ normalized according to \cite{blazek2011} and the other $c$ coefficients set to vanish at $\zf$.}\label{fig:config}
\end{center}
\end{figure}

\begin{figure}
\begin{center}
\includegraphics[width=\textwidth]{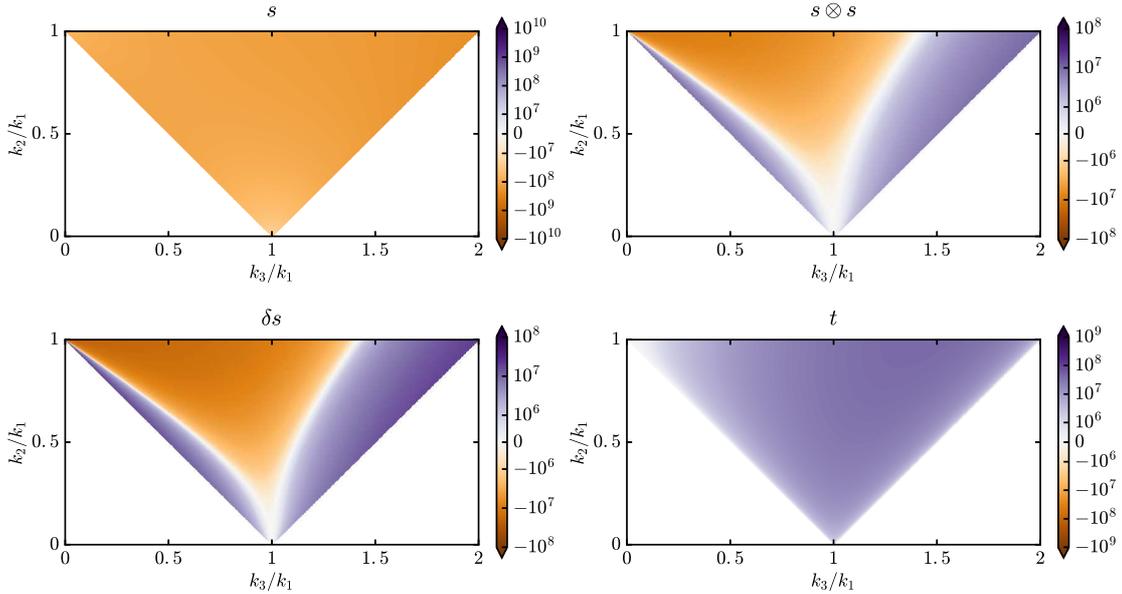}
\caption{The contributions to the E-mode configuration-space \ggi bispectrum in units of $(\mathrm{Mpc}/h)^6$ at $k_1 = 0.1 \, \mathrm{Mpc}/h$ (top left panel of Figure~\ref{fig:config}) from the four terms ($s$, $\sxs$, $\d s$, and $t$). The degeneracy of the terms $c_{\sxs}$ and $c_{\d s}$ for plane-of-sky triangles is visually apparent.}\label{fig:terms}
\end{center}
\end{figure}

\begin{figure}
\begin{center}
\includegraphics[width=0.7\textwidth]{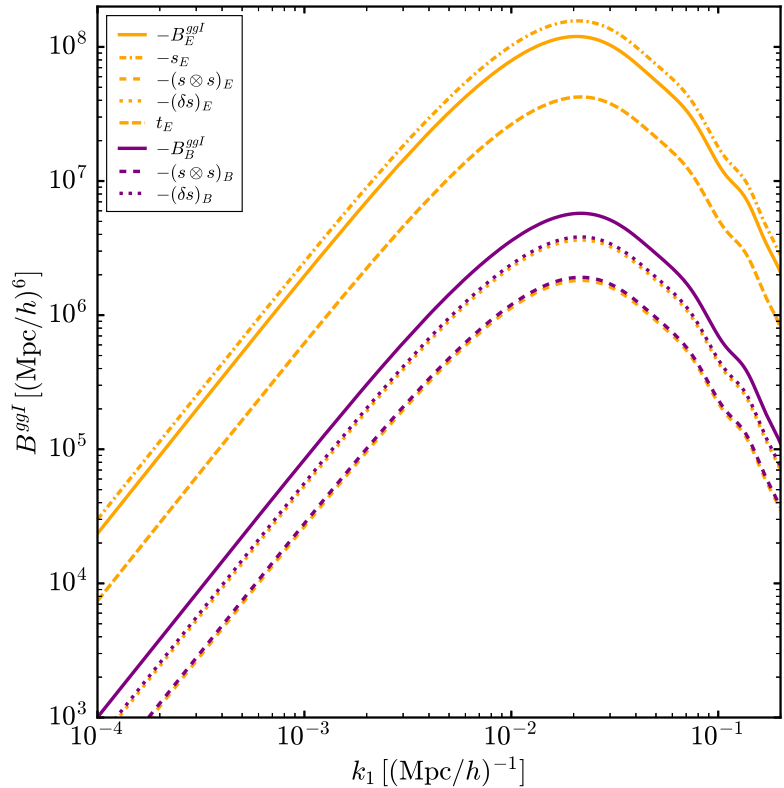}
\caption{Absolute values of the $E$-mode and $B$-mode \ggi bispectra, and their constituent components depending on the $c$ coefficients, as a function of scale $k_1$ for the triangle shape $k_2 = 0.5 k_1$ and $k_3 = k_1$. All quantities except the $t_E$ component are negative for this configuration and therefore have an overall negative sign. Note that $s$ and $t$ do not have $B$-mode components. As in Figures~\ref{fig:config}~and~\ref{fig:terms}, these bispectra were computed assuming the fiducial combined BOSS model with $\zf = 0.49$, $\zi = 6$, and $c_s$ normalized according to \cite{blazek2011} and the other $c$ coefficients set to vanish at $\zi$.}\label{fig:1d}
\end{center}
\end{figure}

Here the \ggi bispectrum is numerically evaluated for our fiducial assumptions. In Figures~\ref{fig:config}~--~\ref{fig:1d}, we adopt the values $\zf = 0.49$ and $\zi = 6$. The fiducial model for these figures is obtained by assuming that only $c_s$ is non-zero at $\zi$, so that the values of these other terms at $\zf$ are completely determined by propagating $c_s(\zi)$ forward in time via the time evolution matrix introduced in Section~\ref{subsec:formalism}. The values of the coefficients at redshift $z = 0.49$ determined in this way are
$c_s =  0.0422$,
$c_{\sxs} = 0.0217$,
$c_{\d s} = -0.0216$, and $c_t = 0.0809$.

Figure~\ref{fig:config} shows the $E$-mode and $B$-mode \ggi bispectra in configuration space for $k_1 = 0.1 \, \mathrm{Mpc}/h$ and $k_1 = 0.01 \, \mathrm{Mpc}/h$. Figure~\ref{fig:terms} shows the contributions to the E-mode bispectrum at $k_1 = 0.1 \, \mathrm{Mpc}/h$ from the four terms individually: $s$, $\sxs$, $\d s$, and $t$. The positive\footnote{Since $c_s$ gives a negative contribution to the bispectrum, but $c_t$ gives a positive contribution, observations (which constrain the sum) will lead to a positive correlation coefficient.} correlation coefficient between the coefficients $c_s$ and $c_t$, which will be presented in Section~\ref{sec:error}, as well as the degeneracy between $c_{\sxs}$ and $c_{\d s}$, are visually evident in this figure. Figure~\ref{fig:1d} shows the scale dependence of the $E$-mode and $B$-mode bispectra for the triangle shape defined by $k_2 = 0.5 k_1$ and $k_3 = k_1$.

\subsection{Time evolution and nonlocality}
\label{subsec:nonlocal}

The $t_{ij}$ term in the expression for $\shij$ is of particular interest because it is nonlocal. That is, $t(\x,z)$ cannot be determined by a local measurement of the gravity gradient $\nabla_i\nabla_j\Phi$ by an observer at $(\x,z)$. This is in contrast to the other terms in the second-order decomposition ($s$, $\d s$, and $\sxs$), which are all local. This fact is significant because under the assumption that the physical processes involved in galaxy formation are local and respond instantaneously to the tidal field, it follows that $c_t = 0$ at the time at which IA is determined by the LSS. Thus any finite value of $c_t$ at lower redshifts must be entirely due to the effects of time evolution, or to the galaxy's ``memory'' of the past history of the tidal field.

This fact suggests that observations of $c_t$ may be used as a probe of galaxy formation processes. First, let $\zia$ denote the formation redshift, \emph{i.e.}, the redshift at which we assume (in the passive evolution model) that IA is determined. Note that $\zia$ is distinct from $\zi$ because we have used $\zi$ to denote simply the initial step in a generic time evolution transformation $T(\zi \rightarrow \zf)$, whereas $\zia$ denotes a \emph{specific} redshift which can be measured in the manner outlined here. If $\sh$ correlations can be observed with enough precision to constrain the individual $c$ coefficients, the value of $c_t$ can be used as an indicator of $\zia$ because it evolves in a deterministic fashion with redshift. A measured value of $c_t(\zf)$ can be traced backwards to estimate $\zia$. Although it is likely overly simplistic to assume that $\zia$ is a well-defined redshift after which IA passively evolves, this method can be used to estimate an ``effective'' or ``average'' value of $\zia$.

It is instructive to consider how to interpret such an effective value. Consider an observational survey of a sample characterized by the probability density function $P_{\rm{sur}}(\zf)$ over a range in observed redshift $\zf$. Suppose the survey samples a galaxy population (\emph{e.g.}, LRGs) with a number density $n_g(z)$ at redshift $z$, and let $dn_g/dz$ denote the associated volumetric formation rate density (in units (Mpc$/h$)$^{-3}$) as a function of redshift. The observed value of $c_t$ in this survey at effective redshift $z\eff$ corresponds to the theoretical quantity given by \begin{equation} \bar{c}_t(z\eff) = \iint c_t(\zf; \zi) \f{1}{n_g(\zf)} \f{dn_g}{dz} \Big|_{\zi} P_{\rm{sur}}(\zf) \, d\zf \, d\zi  \end{equation} where $c_t(\zf;\zi)$ is the value of $c_t$ at final redshift $\zf$ given an initial redshift $\zi$ (\emph{i.e.}, assuming $c_t(\zi) = 0$). Then $\zia$ is obtained by inverting the expression for $c_t(\zf;\zi)$ with respect to $\zi$ and plugging in $\bar{c}_t(\zf)$. In other words, $\zia$ is the inverse of a weighted average of $c_t$ over redshift, where the weighting function is the galaxy formation rate.

\begin{figure}
\begin{center}
\includegraphics[width=0.7\textwidth]{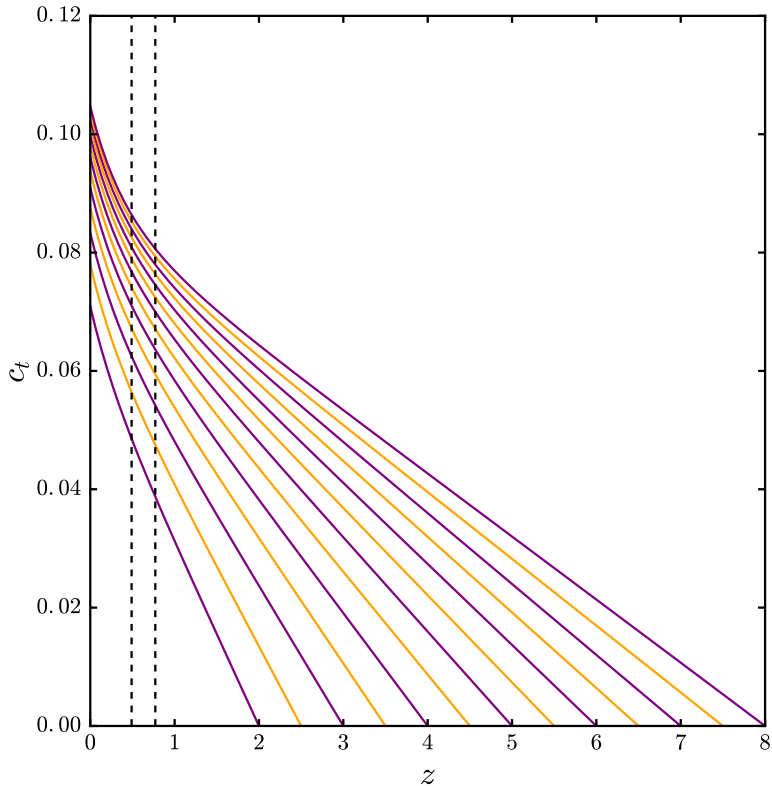}
\caption{The evolution of $c_t$ with $z$ for varying values of $\zia$ from 2 to 8, assuming that $c_t(\zia) = 0$ and that IA passively evolves. The black dashed lines are located at $z = 0.49$ and $z = 0.771$, the effective redshifts of the combined BOSS sample and the DESI LRG sample, respectively.}\label{fig:ct}
\end{center}
\end{figure}

Figure~\ref{fig:ct} shows $c_t$ as a function of $z$ for various assumed values of $\zia$ from $2.0$ to $8.0$. The differences in these functions at $\zf$ are small, but still potentially distinguishable. Forecasts for constraining $c_t$ and the consequent constraints on $\zia$ are discussed in Section~\ref{sec:error}. Note that the differences in the time evolution curves become larger for larger values of $\zf$, as the observation time moves closer to the formation time. Evidently, $c_t$ is a more powerful probe of galaxy formation physics when it is observed at higher redshifts.

\section{Measurement forecasts}
\label{sec:error}

In this section, we use Fisher information matrix analysis to evaluate the feasibility of measuring the non-locality coefficient $c_t$ from the bispectrum in SDSS-BOSS and DESI LRG. We assume that analysis is restricted to triangles with normal vectors $\hat{\bf n}$ satisfying $\theta_{\hat{\bf n},\hat{\bf e}_z} < \pi/6$, \emph{i.e.}, triangles near the plane of the sky\footnote{We have chosen the value $\theta = \pi/6$ such that a given galaxy is subject to an inclination correction of no greater than $1 - \cos^2\theta = 0.25$.}, so that the plane-of-sky approximation $\k\perp\hat{\bf e}_z$ is valid. We assume a known cosmology and forecast the measurement precision on the IA parameters alone. The derivation of the analytic expression for the Fisher information is presented in Appendix~\ref{apx:fisher}. We find that the matrix element $\mathcal I_{mn}$ is given by \begin{align}
\mathcal I_{mn} &= \f{Vf_\tri}{8 \pi^4} \int_{k_1 < k_2} k_1 \, k_2 \, k_3 \, \f{\dd}{\dd c_m} B (\k_1, \k_2, -(\k_1 + \k_2)) \f{\dd}{\dd c_n} B (\k_1, \k_2, -(\k_1 + \k_2)) \nonumber \\
&\times \lp b_1(z)^2 D(z)^2 \Plin(k_1) + \f{1}{\nbar} \rp^{-1} \lp b_1(z)^2 D(z)^2 \Plin(k_2) + \f{1}{\nbar} \rp^{-1} \nonumber \\&\times \lp c_s(z)^2 D(z)^2 (1+z)^2 f_{EB}(k_3^{\alpha})^2 \Plin(k_3^{\alpha}) + \f{\sgam^2}{\nbar} \rp \mathrm d k_1 \, \mathrm d k_2 \, \mathrm d k_3.
\end{align}

\subsection{SDSS-BOSS}
\label{subsec:boss}

We can now numerically evaluate the Fisher information matrix using survey parameters from SDSS-BOSS DR12 \cite{reid2016} as in Section~\ref{sec:num}. As mentioned in Section~\ref{sec:num}, the survey consists of two subsamples, CMASS ($N_{\mathrm{gal}} = 777202$, $V_{\mathrm{eff}} = 1.70 \, (\mathrm{Gpc}/h)^3$, $z\eff = 0.57$), and LOWZ ($N_{\mathrm{gal}} = 361762$, $V_{\mathrm{eff}} = 0.766 \, (\mathrm{Gpc}/h)^3$, $z\eff = 0.32$). We combine the samples into a single BOSS sample ($N_{\mathrm{gal}} = 1138964$, $V_{\mathrm{eff}} = 2.47 \, (\mathrm{Gpc}/h)^3$, $z\eff = 0.49$) and compute its Fisher information in each of the E and B modes, then combine the resulting two matrices. We impose a cutoff of the integrals at $k_1,k_2,k_3<0.2h\,{\rm Mpc}^{-1}$ in order to reject the deeply non-linear scales where tree-level calculations fail. (Choosing a cutoff of $0.1h\,{\rm Mpc}^{-1}$ instead increases our derived errorbars by a factor of $\sim 4$.) We adopt the value $\sgam = 0.170$, the measured root-mean-square ellipticity per component for this sample \cite{eisenstein2011, dawson2013, gunn2006, alam2015, smee2013, gunn1998, doi2010}.\footnote{Specifically, we took the $(a-b)/(a+b)$ ellipticities from the de Vaucouleurs fits used in \cite{martens2018}. With this ellipticity convention, the response factor converting from mean ellipticity to shear is unity.} Note that the measured RMS ellipticity includes the contributions from both the intrinsic shape noise of the BOSS galaxies and from measurement noise in SDSS. From the Fisher information matrices, we forecast the error bars and correlation coefficients of the bias parameters, as shown in Table~\ref{table:boss-params}. The fiducial parameters shown in the table are consistent with those in Section~\ref{subsec:num_bispec}.

All of these quantities are sufficiently different from zero that they are potentially detectable even in existing datasets. It should be noted that time evolution due to advection ensures that the coefficients can all be nonzero \emph{even if one assumes a very simple model for galaxy formation} (\emph{e.g.}, strictly linear alignment at the time of formation, such as we have assumed in this work). In other words, all coefficients except for $c_s$ can vanish at the time of galaxy formation and still be nonzero -- and in fact statistically different from zero -- by the time that observations are made.

\begin{table} \centering
\begin{tabular}{c|r|rrr}
Parameters\ $c_n$ & $c_n\pm\sigma(c_n)$~~~~ & \multicolumn3c{Correlation coefs.\ $\rho_{mn}$}
\\
\hline
$c_s$ & $0.0422 \pm 0.0007$ & $1.00$ & $-0.17$ & $0.73$ \\
$ c_{\d s} - \tfrac13(1+z) c_{\sxs}$ & $-0.0324 \pm 0.0040$ & $-0.17$ & $1.00$ & $-0.17$ \\
$c_t$ & $0.0809 \pm 0.0080$ & $0.73$ & $-0.17$ & $1.00$
\end{tabular}
\caption{Fiducial parameter values, uncertainties estimated from the Fisher matrix (\emph{i.e.}, marginalized over the remaining parameters), and correlation coefficients for the combined BOSS CMASS + LOWZ sample ($z=0.49$). This matrix includes constraints from both $E$ and $B$ modes.}
\label{table:boss-params}
\end{table}

\subsection{DESI LRG}
\label{subsec:desi-lrg}

We shall also produce a forecast for the upcoming DESI survey, which will provide the largest planned sample of LRGs. We compute the expected total galaxy number, effective volume, and effective redshift of the planned DESI luminous red galaxy (LRG) sample from information provided in \cite{desicollaboration2016}. In particular, we compute the effective volume in the manner of Eq.~50 of \cite{reid2016}. As we did for the BOSS forecast, we compute E- and B-mode information separately for the DESI LRG sample ($N_{\mathrm{gal}} = 3948000$, $V_{\mathrm{eff}} = 6.76 \, (\mathrm{Gpc}/h)^3$, $z\eff = 0.771$) and then combine the resulting matrices. The resulting uncertainties on the fiducial parameters are shown in Table~\ref{table:desi-params}. Note that the fiducial parameter values in Table~\ref{table:desi-params} are different from those in Table~\ref{table:boss-params}, because the effective redshift of DESI LRG will be different from that of BOSS. We adopt the shape noise value per component $\sigma_{\gamma} = 0.186$, calculated from Gaussian fitting of the ellipticity distribution of the DECaLS\footnote{http://legacysurvey.org/decamls/} DR3 magnitude-cut sample with an additional DESI-like (see \cite{desicollaboration2016}) color cut of $r-z > 1.5$ applied.

\begin{table} \centering
\begin{tabular}{c|r|rrr}
Parameters\ $c_n$ & $c_n\pm\sigma(c_n)$~~~~ & \multicolumn3c{Correlation coefs.\ $\rho_{mn}$}
\\
\hline
$c_s$ & $0.0408 \pm 0.0004$ & $1.00$ & $-0.16$ & $0.73$ \\
$c_{\d s} - \tfrac13(1+z) c_{\sxs}$ & $-0.0298 \pm 0.0025$ & $-0.16$ & $1.00$ & $-0.16$ \\
$c_t$ & $0.0745 \pm 0.0051$ & $0.73$ & $-0.16$ & $1.00$
\end{tabular}
\caption{Fiducial parameter values, uncertainties, and correlation coefficients for the planned DESI LRG sample $(z = 0.771)$. This matrix includes constraints from both $E$ and $B$ modes.}
\label{table:desi-params}
\end{table}

\subsection{Constraints on $\zia$}
\label{subsec:zi}

\begin{figure}
\begin{center}
\includegraphics[width=0.7\textwidth]{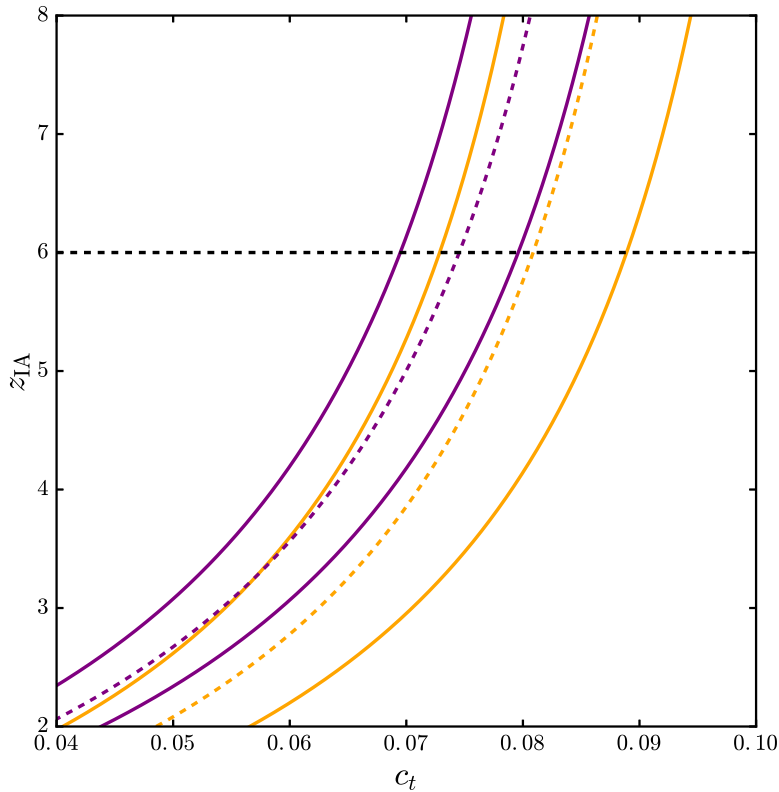}
\caption{The estimated effective formation redshift $\zia$ as a function of the measured value of $c_t$ in BOSS (orange dashed line) and DESI LRG (purple dashed line). Solid lines are $1\sigma$ error contours. The black dashed line marks the fiducial value $\zia = 6.0$ adopted for the calculations of the quantities shown in Figures~\ref{fig:config}--\ref{fig:1d} and Tables~\ref{table:boss-params} and \ref{table:desi-params}. The offset between the two central curves is due to the fact that the BOSS and DESI LRG surveys have different effective (observation) redshifts, so we would expect different observed values of $c_t$ for the same value of $\zia$. }\label{fig:zia}
\end{center}
\end{figure}

Potential constraints on the effective formation redshift are also of interest. Figure~\ref{fig:ct} demonstrates how $\zia$ is constrained by a measurement of $c_t$. Because $c_t$ should vanish at $\zia$, its value at the observed redshift, $c_t(\zf)$, is entirely determined by evolving $c_s$ forward in time via the off-diagonal matrix element $T_{t,s}$. Given a measured value of $c_t$, we can reverse this process to determine $\zia$. The resulting constraint on $\zia$ is shown in Figure~\ref{fig:zia}, with errorbars based on based on the $c_t$ errorbars from the Fisher information matrix.\footnote{Note that the $c_t$ errorbars are marginalized over the other two parameters. Actually, if $c_s$ is already well known from observations such as \protect{\cite{blazek2011}}, marginalizing over it is not strictly necessary, but the forecasted errorbars on $c_s$ in this work are small enough that this likely does not have a large effect on the constraints in Figure~\ref{fig:zia}.} The constraint is tighter for relatively low $\zia$ and less tight at higher redshifts. We see also that the constraining power of DESI is greater than that of BOSS.

\section{Discussion}
\label{sec:disc}

In this work, we have obtained several results on the time evolution of intrinsic alignments. In Sections~\ref{subsec:formalism}~and~\ref{subsec:matrix}, we have written an IA expansion, complete to second order in $\dlin$, and determined how the associated IA bias coefficients evolve in time according to passive evolution and advection. The transformation matrix elements for this evolution are presented in Section~\ref{subsec:adv}. We have then used this result to choose reasonable fiducial values at $\zf = 0.57$ for these coefficients (which have not yet been measured), and present the tree-level density-density-IA bispectrum assuming these fiducial values in Sections~\ref{subsec:bispec}~and~\ref{subsec:num_bispec}. Because advection in particular presents a unique probe of galaxy formation physics, we have demonstrated in Section~\ref{subsec:nonlocal} the utility of the $c_t$ coefficient for studying IA in this context. Finally, in Sections~\ref{subsec:boss}~and~\ref{subsec:desi-lrg}, we have presented Fisher information analysis demonstrating the constraining power of SDSS-BOSS and DESI for both the IA coefficients themselves and (in Section~\ref{subsec:zi}) the galaxy physics information encoded in them.

To second order in $\dlin$, the bias coefficients describing IA evolve in time even under the assumption of passive evolution, due to nonlinear growth of structure and the mixing of bias coefficients due to galaxy advection.
This passive evolution assumption is often adopted in theoretical work on IA, although the extent to which it is correct remains an open question. Future observations could either verify this assumption or call it into question, depending on whether or not the IA coefficients are observed to transform under time evolution in a manner consistent with the calculations in this work.

The advection effect arises because galaxies have peculiar velocities which cause their comoving positions to change over time. Notably, because the local gravitational field at a point in spacetime determines the acceleration of a galaxy at that point, a galaxy's peculiar velocity will retain the memory of the gravitational effects that the galaxy has felt throughout its history. Advection therefore results in a component of $\sh$ with nonlocal dependence on the cosmological fields. The nonlocal component is of particular interest as a probe of galaxy formation physics, because it can potentially be used to trace IA evolution back to the time of galaxy formation and determine the approximate redshift at which the IA were set. We therefore suggest that in addition to cosmology, galaxy formation physics is a compelling motivation to attempt a measurement of $c_t$ in future IA surveys. In addition, better constraints on the effective redshift of IA determination can be obtained from measurements of $c_t$ at higher redshifts, a fact which should motivate efforts to push IA observations beyond $z \sim 0.5$. The relative amplitudes of the $c$ coefficients and their time evolution could also provide a probe of cosmology and modified gravity, since they are determined by the growth of structure.

We noted in Section~\ref{sec:error} that one consequence of the effect of advection on IA statistics is that a galaxy population can exhibit non-vanishing values of all three second-order IA bias coefficients even if only the linear (tidal alignment) coefficient was non-zero at the time at which IA was first determined. This is a salient point because it implies that higher-order terms should not be neglected even for galaxy populations that are assumed to have a very simple functional dependence of IA on the LSS. For example, one should not assume that the so-called ``tidal torquing term'' ($\sxs$) is only relevant for rotationally-supported blue galaxies. Time evolution allows for such a term to arise in any galaxy sample through mechanisms quite distinct from actual tidal torquing.

In this work, we have presented the first calculation of the \ggi bispectrum including all of the terms that arise at second order in SPT. In forthcoming work, we shall extend this method to a complete description of IA at third order in SPT, which will be used to model the density-IA power spectrum to one-loop order. In addition to allowing this analysis to be applied to two-point statistics, a higher-order calculation of the time evolution of IA will allow for more robust constraints on galaxy formation using multiple bias coefficients. That work will also address the question of whether the transformation of IA coefficients we found at second-order can be generally extended to higher order, analogous to how we demonstrated that the linear order transformation of the $c_s$ coefficient was the same when extended to second order. In another future work, we shall address the application of IA time evolution as a probe of modified gravity.

\appendix
\section{Decomposition of the convolution integrals}
\label{apx:decom}

In this appendix we present in more detail the decomposition of the advection term into the four components of the IA field. For convenience, we define the variables $\qp$ and $\mu$ by $\qp \equiv \k - \q$ and $\mu \equiv \q \cdot \qp / qq'	$. We also define a convolution operator as follows: \begin{equation} \conv f(\q, \qp) \equiv \int f(\q, \qp) \dlin(\q) \dlin(\qp) \fdq. \end{equation} The advection term, up to a function of $z$, is given by \begin{equation} \label{ae:adv} (\nabla s)_{ij} \cdot (\xi - \xf) = -\conv \f{\q \cdot \qp}{q'^2} \lp \f{q_i q_j}{q^2} - \f 13 \dk_{ij} \rp. \end{equation} 

\noindent Similarly, the second-order components can be written as
\begin{align}
s_{ij}^{(2)} &= -\conv \lp \f{k_i k_j}{k^2} - \f 13 \dk_{ij} \rp \lp \f 57 + \f 27 \mu^2 + \f{\q \cdot \qp}{q'^2} \rp,
\\ (\sxs)_{ij} &= \conv \lp \lp \f{q_i q_k}{q^2} - \f 13 \dk_{ik} \rp \lp \f{q'_j q'_k}{q'^2} - \f 13 \dk_{jk} \rp - \f 13 \dk_{ij} \lp \mu^2 - \f 13 \rp \rp,
\\ \lp \d s \rp_{ij} &= -\conv \lp \f{q_i q_j}{q^2} - \f 13 \dk_{ij} \rp, {\rm ~~and}
\\ \label{ae:t} t_{ij} &= -\conv \f 27 \lp \f{k_i k_j}{k^2} - \f 13 \dk_{ij} \rp \lp \mu^2 - 1 \rp .
\end{align} 

\noindent We begin by symmetrizing
Eqs.~\ref{ae:adv}--\ref{ae:t} in $\q$ and $\qp$, as follows: \begin{align}
\lp(\nabla s)_{ij} \cdot (\xi - \xf)\rp_{\sym} &= -\conv \f 12 (\q \cdot \qp)   \lp \f{q_i q_j + q'_i q'_j}{q^2q'^2} - \f 13 \lp \f{1}{q^2} + \f{1}{q'^2} \rp\dk_{ij} \rp, \\
s_{ij, \sym}^{(2)} &= -\conv \lp \f{k_i k_j}{k^2} - \f 13 \dk_{ij} \rp \lp \f 57 + \f 27 \mu^2 + \f 12 (\q \cdot \qp) \lp\f{1}{q^2} + \f{1}{q'^2} \rp \rp, \\
(\sxs)_{ij, \sym} &= \conv \bigg( \f 12 \lp \f{q_i q_k}{q^2} - \f 13 \dk_{ik} \rp \lp \f{q'_j q'_k}{q'^2} - \f 13 \dk_{jk} \rp \nonumber \\
&~~~~+ \f 12 \lp \f{q_j q_k}{q^2} - \f 13 \dk_{jk} \rp \lp \f{q'_i q'_k}{q'^2} - \f 13 \dk_{ik} \rp - \f 13 \dk_{ij} \lp \mu^2 - \f 13 \rp \bigg),\\
\lp\d s\rp_{ij, \sym} &= -\conv \lp \f 12 \lp \f{q_i q_j}{q^2} + \f{q'_i q'_j}{q'^2} \rp- \f 13 \dk_{ij} \rp, ~~{\rm and}
\\ t_{ij, \sym} &= -\conv \f 27 \lp \f{k_i k_j}{k^2} - \f 13 \dk_{ij} \rp \lp \mu^2 - 1 \rp. 
\end{align}

\noindent We shall now verify that the advection expression may be decomposed in terms of these components. Omitting some subscripts for convenience, we first note that \begin{equation} \sxs - \f 23 \d s = \conv \lp \f 12 (\q \cdot \qp) \f{q_i q'_j + q'_i q_j}{q^2 q'^2} - \f 13 \mu^2 \dk_{ij} \rp. \end{equation} 

\noindent Furthermore, \begin{equation} -\nabla s \cdot \d \x + \sxs - \f 23 \d s = \conv \lp \f 12 \f{\q \cdot \qp}{q^2 q'^2} k_i k_j - \f 13 \dk_{ij} (\q \cdot \qp) \lp \f{1}{q^2} + \f{1}{q'^2} \rp - \f 13 \mu^2 \dk_{ij} \rp. \end{equation}

\noindent With the application of some algebra, the reader may verify that \begin{equation} -\nabla s \cdot \d \x + s^{(2)} + \sxs - \f 23 \d s = \conv \f 57 \lp \f{k_i k_j}{k^2} - \f 13 \dk_{ij} \rp \lp \mu^2 - 1 \rp \end{equation}

\noindent and therefore \begin{equation} \nabla s \cdot \d \x = s^{(2)} + \sxs - \f 23 \d s + \f 52 t. \end{equation}

This calculation provides the signs and multiplicative constants that appear in the time evolution matrix elements (Eqs.~\ref{eq:matr-first}--\ref{eq:matr-last}). The functional dependence of the matrix elements on redshift is determined by the derivation in Section~\ref{subsec:adv} and by the time evolution of the cosmological fields in SPT. In particular, the functional forms of Eqs.~\ref{eq:matr-first}~--~\ref{eq:matr-last-diag} are simply given by \begin{equation} T_{f,f} = \f{f(\zi)}{f(\zf)} \end{equation} while the functional forms of Eqs.~\ref{eq:matr-first-odiag}~--~\ref{eq:matr-last} come from the redshift-dependent part of Eq.~\ref{eq:adv}.

\section{Lagrangian derivation and basis completeness}
\label{apx:lag}

This section is motivated by the question of whether $s_{ij}$, $(s\otimes s)_{ij}$, $\delta s_{ij}$, and $t_{ij}$ form a complete basis at second order, or whether there are additional terms that arise at the same order. (Note that in this discussion, we neglect higher-derivative terms at a given order which reflect physics at the scale of the halo or galaxy including smoothing of the relevant cosmological fields (see, \emph{e.g.}, \cite{angulo2015, blazek2017}). These terms are suppressed on large scales but become important when modeling scales that are of the same order as the halos/galaxies.) After all, some highly non-obvious combinations of fields may obey the relevant symmetry properties, \emph{e.g.}, the $\psi$ field in Eq.~8 of \cite{mcdonald&roy2009}. As it happens, the completeness of the basis is manifest when treated in the Lagrangian formalism. (For an exposition of Lagrangian perturbation theory, see for example \cite{tatekawa2004}.)

We recall that -- so long as we use the $F_n$ and $G_n$ perturbation theory kernels from the EdS cosmology, which turns out to be a good approximation even in $\Lambda$CDM -- the density and tidal field experienced by an individual particle can be written as a Taylor series in $D(t)$ up through $n$th order (where $n$ is the order of perturbation theory considered). Furthermore, the velocity gradient can be written entirely in terms of the history of the tidal field and the density.\footnote{This is in fact true in full general relativity, and follows from Eq.~4.25 of \protect{\cite{hawking&ellis1973}}: for geodesics (dark matter particle trajectories), the acceleration $\dot V_\alpha=0$. Moreover, for scalar initial conditions, $\omega_{\alpha\beta}=0$ and will remain zero for all time. Then the 6 components of the symmetric velocity gradient $\theta_{\alpha\beta}$ obey a first-order system of coupled ordinary differential equations, with initial conditions in the early Universe that $\theta_{\alpha\beta} = H\delta_{\alpha\beta}$ and with a source that is a $3\times 3$ symmetric part of the Riemann tensor component $R_{\alpha 4\beta 4}$. The trace of this is proportional to the matter density, and the traceless-symmetric part is the tidal field. Thus the full history of the matter density and the tidal field must determine the full history of the velocity gradient.} Therefore, all information on the history of the density and tidal field through order $n$ accessible to a galaxy is captured in $\d$, $\ld\d/\ld t$, \ldots, $\ld^{n-1}\d/\ld t^{n-1}$ and $s_{ij}$, $\ld s_{ij}/\ld t$, \ldots, $\ld^{n-1}s_{ij}/\ld t^{n-1}$, where \begin{equation} \f{\ld}{\ld t} = \frac{\dd}{\dd t} + (1+z) u_k \nabla_k \label{eq:lagr} \end{equation} (where $u_k \equiv av_k$) is a Lagrangian derivative. Thus in our case ($n=2$), we have available two scalars and two traceless-symmetric tensors $\d$, $\ld \d/ \ld t$, $s_{ij}$ and $\ld s_{ij}/\ld t$. As has been done for the galaxy case, it is convenient to make a change of basis so that some of the quantities appear at first-order in perturbation theory, and some appear at second-order:
\begin{eqnarray}
1^{\rm st}+\,{\rm order}:&& ~~ \d, ~~s_{ij},
\nonumber \\
2^{\rm nd}+\,{\rm order}:&& ~~ H^{-1}\frac{\ld \d}{\ld t} - \d, ~~
H^{-1}\frac{\ld s_{ij}}{\ld t}.
\label{eq:12}
\end{eqnarray}
where we took $\ld\ln D/ \ld\ln a=1$ for consistency with the EdS kernels. Now we search for quantities that are overall first or second order -- \emph{i.e.}, are \emph{(i)} 1$^{\rm st}$ order quantities; \emph{(ii)} products of two 1$^{\rm st}$ order quantities, or \emph{(iii)} 2$^{\rm nd}$ order quantities in Eq.~\ref{eq:12} -- and are traceless-symmetric tensors. From the finite list of quantities in Eq.~\ref{eq:12} we may enumerate all possibilities: there is \emph{(i)} $s_{ij}$; \emph{(ii)} $\d s_{ij}$ and $(s\otimes s)_{ij}$; and \emph{(iii)} $H^{-1}\ld s_{ij}/\ld t$. 

We shall now see how a Lagrangian treatment of the time evolution of $\sh$ makes the $(s, \sxs, \d s, t)$ basis manifestly complete. The passive evolution model (\emph{cf.} Eq.~\ref{eq:pe}) implies that $\ld \sh / \ld t = 0$ and therefore \begin{equation} 0 = \sum_{n = s, \sxs, \d s, t} \lp n_{ij}(\ti) \f{dc_n}{dt} + c_n \f{\ld n_{ij}}{\ld t} \rp = \sum_n \lp \frac{\dd}{\dd t}(c_n n_{ij}) + c_n v_k \nabla_k n_{ij} \rp. \end{equation} Note that the advective portion of the Lagrangian derivative will vanish at second order in PT for $n \neq s$. We shall now Taylor expand the expression for $\sh$ to first order in $t$: \begin{align} \shij(\xf, \tf) = \shij(\xi, \ti) \nonumber &= \shij(\xf, \ti) + c_s v_k \nabla_k s_{ij} \Big|_{\xf, \ti}(\ti - \tf) + \ldots \\ &= \shij(\xf, \zi) + c_s(\zi) \nabla_k s_{ij}(\xf, \zi) \cdot (\xi - \xf)_k + \ldots. \label{eq:lagr-result}\end{align} We have recovered Eqs.~(\ref{eq:pe})~and~(\ref{eq:gij}). 

The point of this exercise has been to show that the basis given in Eq.~\ref{eq:12} is equivalent to the enumeration of terms in Eqs.~(\ref{eq:tidal}--\ref{eq:t}). We thus conclude that the latter is a complete basis. Indeed, the equivalence of the Lagrangian and Eulerian formalisms implies that, following the calculation in Appendix~\ref{apx:decom},
\begin{equation}\label{eq:basis}
H^{-1}\frac{\ld s_{ij}}{\ld t} = -\f{(s\otimes s)_{ij}}{(1+z)} + \frac23\d s_{ij} - \frac52t_{ij}.
\end{equation}

\section{Group theory of the $\sxs$ and $\d s$ terms}
\label{apx:group}

In this section, we present a group-theoretic argument for the degeneracy of the $\sxs$ and $\d s$ terms for plane-of-sky triangles.

Recall that the group O(2) of (proper and improper) rotations has a 2-dimensional spin-$m$ representation for each $m=1,2,3,\ldots$, and for $m=0$ there are two 1-dimensional representations $0^+$ and $0^-$ that are even and odd respectively under reflections through a plane containing the line of sight. For wave vectors in the plane of the sky, $s$ has components that transform under the $0^+\oplus 2$ representation of the O(2) group representing rotations around the line of sight. (Although a general traceless-symmetric tensor has 5 components and transforms under $0^+\oplus1\oplus 2$, the two $m=1$ components ($s_{13},s_{23}$) are zero for wave vectors in the plane of the sky.) By contrast, the density $\delta$ transforms under the $0^+$ representation, and is simply proportional to the $0^+$ component of $s$ (the component $s_{33}$ that is invariant under rotations around the line-of-sight is proportional to $\delta$: $s_{33} = \tfrac13(1+z)\delta$). Now we see that $\d s$ transforms under $0^+\otimes(0^+\oplus 2) = 0^+\oplus 2$, and thus has one part that transforms under the spin 2 representation of O(2). By contrast, $\sxs$ has parts that transform under the symmetric tensor product of $0^+\oplus 2$ with itself, which is $0^+ \oplus 0^+ \oplus 2 \oplus 4$.\footnote{Generally, $(0^+\oplus 2)\otimes (0^+\oplus 2) = 0^+ \oplus0^+ \oplus 0^- \oplus 2 \oplus 2 \oplus 4$, which has dimension $3^2=9$; however only $3(3+1)/2=6$ of these are part of the symmetric tensor product.} The only spin 2 part in the decomposition of $\sxs$ arises from multiplying the spin $0^+$ and spin 2 parts of $s$, and is thus proportional to the spin 2 part of $\d s$. It is therefore intuitive that -- using only modes with wave vectors in the plane of the sky -- the bias coefficients for $\d s$ and $\sxs$ would be degenerate.

\section{Analytic form of the Fisher information matrix}
\label{apx:fisher}

In this appendix we present the calculation of the Fisher information matrix, which is numerically evaluated in Section~\ref{sec:error}. (For an exposition of the use of Fisher information in cosmology, see for example \cite{heavens2009}.)

The Fisher information matrix elements are given by \begin{equation}\mathcal I_{mn} = \sum_{\mu\nu} \f{\dd}{\dd c_m} B (\tri^{\mu}) \f{\dd}{\dd c_n} B (\tri^{\nu}) \lp \cov (B,B)^{-1} \rp_{\mu \nu},\label{eq:fisher} \end{equation} 
where the symbol $\tri$ denotes the ordered triple $(\k_1, \k_2, \k_3)$ with $\k_1+\k_2+\k_3=0$, and the Greek indices $\mu$ and $\nu$ denote which triangle is being considered. The partial derivatives $\dd B / \dd c_s$, $\dd B / \dd c_{\sxs}$, $\dd B / \dd c_{\d s}$, and $\dd B / \dd c_s$ are readily obtained from Eqs.~\ref{eq:ebispec}~and~\ref{eq:bbispec}. The covariance matrix calculation can be simplified by assuming Gaussianity and by noting that for all $k \gtrsim 0.05 \, (\mathrm{Mpc}/h)^{-1}$ the shape noise dominates over the intrinsic alignments:\footnote{The range of $k$ in which this inequality is invalid contributes very little to the total Fisher information. Computing the Fisher matrix with all $k < 0.05 \, (\mathrm{Mpc}/h)^{-1}$ excised gives identical results (to within the quoted precision) for the forecasted errorbars in Tables~\ref{table:boss-params}~and~\ref{table:desi-params}.}
\begin{equation}
\label{eq:sgam-cond}
\frac{\sgam^2}{\nbar} \gg c_s(\zf)^2 (1+\zf)^2 D(\zf)^2 \Plin(\k)
\end{equation}
Then we can evaluate the covariance matrix using Wick's theorem:
\begin{align} \la B(\tri^{\alpha}) B(\tri^{\beta}) \ra\conn 
&\approx  \la \d(\k_1^{\alpha}) \d(\k_1^{\beta}) \ra \la \d(\k_2^{\alpha}) \d(\k_2^{\beta}) \ra \la \gamma(\k_3^{\alpha}) \gamma(\k_3^{\beta}) \ra \nonumber \\ 
&+ \label{eq:gauss2} \la \d(\k_1^{\alpha}) \d(\k_2^{\beta}) \ra \la \d(\k_2^{\alpha}) \d(\k_1^{\beta}) \ra \la \gamma(\k_3^{\alpha}) \gamma(\k_3^{\beta}) \ra,
\end{align}
which simplifies at leading order (see, \emph{e.g.}, \cite{kayo2013, takada&jain2004}) to
\begin{align}
\cov (B,B)^{\alpha \beta} &= V \lp \dk_{\k_1^{\alpha}, \k_1^{\beta}} \dk_{\k_2^{\alpha}, \k_2^{\beta}} + \dk_{\k_1^{\alpha}, \k_2^{\beta}} \dk_{\k_2^{\alpha}, \k_1^{\beta}} \rp \dk_{\k_3^{\alpha}, \k_3^{\beta}} \nonumber \\ 
&\times \lp b_1(z)^2 D(z)^2 \Plin(k_1^{\alpha}) + \f{1}{\nbar} \rp \lp b_1(z)^2 D(z)^2 \Plin(k_2^{\alpha}) + \f{1}{\nbar} \rp \nonumber \\ 
&\times \lp c_s(z)^2 D(z)^2 (1+z)^2 f_{EB}(k_3^{\alpha})^2 \Plin(k_3^{\alpha}) + \f{\sgam^2}{\nbar} \rp.
\label{eq:CovBB}
\end{align}
Here we have used Kronecker deltas for the Fourier modes. Note that the density of modes in $\k$-space is $V/(2\pi)^3$, where $V$ is the comoving volume of the survey. Note also that the $\d s$, $\sxs$, and $t$ terms do not contribute to the covariance at this order. (For the $2^{\rm nd}$ order terms to contribute, we would need to calculate the covariance at one-loop order and would therefore require $\sh$ to third order.)

We can now invert the covariance matrix $\cov (B,B)^{\alpha \beta}$ and evaluate Eq.~\ref{eq:fisher}. First, to avoid double-counting of triangles containing the same information, we impose the condition $k_1^\mu<k_2^\mu$. Then we make the replacements
\begin{equation}
\sum_\mu \rightarrow V^2 \int \frac{d^3\k_1^\mu}{(2\pi)^3} \frac{d^3\k_2^\mu}{(2\pi)^3}
~~~{\rm and}~~~
\dk_{\k_1^{\alpha}, \k_1^{\beta}} \rightarrow \frac{(2\pi)^3}{V} \delta^{(3)}(\k_1^\alpha-\k_1^\beta).
\end{equation}
(The latter replacement is only needed for $\k_1$ and $\k_2$, since the Kronecker deltas for $\k_3$ in Eq.~\ref{eq:CovBB} are trivially equal to one if those for $\k_1$ and $\k_2$ are non-zero.) The $\delta$-functions collapse the resulting integrals in Eq.~\ref{eq:fisher} to a 6D integral over one triangle ($\int d^3\k_1^\mu\,d^3\k_2^\mu$):
\begin{align}
\mathcal I_{mn} &= V \int_{k_1 < k_2} \f{\dd}{\dd c_m} B (\k_1, \k_2, -(\k_1 + \k_2)) \f{\dd}{\dd c_n} B (\k_1, \k_2, -(\k_1 + \k_2)) \nonumber \\
&\times \lp b_1(z)^2 D(z)^2 \Plin(k_1) + \f{1}{\nbar} \rp^{-1} \lp b_1(z)^2 D(z)^2 \Plin(k_2) + \f{1}{\nbar} \rp^{-1} \nonumber \\ &\times \lp c_s(z)^2 D(z)^2 (1+z)^2 f_{EB}(k_3^{\alpha})^2 \Plin(k_3^{\alpha}) + \f{\sgam^2}{\nbar} \rp
\frac{d^3\k_1}{(2\pi)^3} \frac{{\mathrm d}^3\k_2}{(2\pi)^3}.
\end{align}

Finally, we make use of the fact that the 6-D integral for the triangle can be converted to an integral over the 3 side lengths ($k_1$, $k_2$, and $k_3$) and the 3 Euler angles describing the orientation of the triangle (the direction of the normal to the triangle, described by longitude $\phi_n$ and co-latitude $\theta_n$; and the position angle $\psi_n$). The conversion is:
\begin{equation}
\frac{d^3\k_1}{(2\pi)^3} \frac{{\mathrm d}^3\k_2}{(2\pi)^3} =
\frac1{8\pi^4} k_1k_2k_3\,{\mathrm d}k_1\,{\mathrm d}k_2\,{\mathrm d}k_3
\, \frac{\sin\theta_n\,{\mathrm d}\theta_n\,{\mathrm d}\phi_n\,{\mathrm d}\psi_n}{8\pi^2},
\end{equation}
where we have normalized the denominator of the angular variables using the fact that the integral of the angular element over all orientations is $\int \sin\theta_n\,{\mathrm d}\theta_n\,{\mathrm d}\phi_n\,{\mathrm d}\psi_n=8\pi^2$ (alternatively, the volume of the group SO(3) is $8\pi^2$). If we integrate over a fraction $f_\tri$ of the orientations of the triangle, then we have:
\begin{align}
\mathcal I_{mn} &= \f{Vf_\tri}{8 \pi^4} \iiint_{k_1 < k_2} k_1 \, k_2 \, k_3 \, \f{\dd}{\dd c_m} B (\k_1, \k_2, -(\k_1 + \k_2)) \f{\dd}{\dd c_n} B (\k_1, \k_2, -(\k_1 + \k_2)) \nonumber \\
&\times \lp b_1(z)^2 D(z)^2 \Plin(k_1) + \f{1}{\nbar} \rp^{-1} \lp b_1(z)^2 D(z)^2 \Plin(k_2) + \f{1}{\nbar} \rp^{-1} \nonumber \\ &\times \lp c_s(z)^2 D(z)^2 (1+z)^2 f_{EB}(k_3^{\alpha})^2 \Plin(k_3^{\alpha}) + \f{\sgam^2}{\nbar} \rp \mathrm d k_1 \, \mathrm d k_2 \, \mathrm d k_3.
\end{align}
Since our calculation is technically valid for ``face-on'' triangle configurations, we should use a value of $f_\tri<1$. A simple cut on the configurations would be to require the triangle normal to be within an angle $\theta_{\rm max}$ of the line of sight, \emph{i.e.},\ $|\cos\theta_n|>\cos\theta_{\rm max}$. This leads to $f_\tri = 1-\cos\theta_{\rm max}$. We have adopted $\theta = \pi/6$, which gives $f_{\tri} = 1 - \sqrt{3}/2$.

\acknowledgments

The authors thank the anonymous referee for helpful comments. DMS acknowledges the support of an NSF Graduate Research Fellowship. CMH is supported by the Simons Foundation, NASA, and the US Department of Energy. JAB acknowledges the support of a Swiss National Science Foundation Ambizione Fellowship. The authors thank D. Martens for providing the version of the BOSS catalog used for a previous project, and O. Kauffmann and H.~Y. Shan for help with the DECaLS catalog. The authors also thank E.~M.~Huff, O.~Dor\'{e}, and P.~F.~Hopkins for helpful discussions and D.~Wittman for the notes available at \texttt{http://wittman.physics.ucdavis.edu/Fisher-matrix-guide.pdf}.

Funding for SDSS-III has been provided by the Alfred P. Sloan Foundation, the Participating Institutions, the National Science Foundation, and the U.S. Department of Energy Office of Science. The SDSS-III web site is http://www.sdss3.org/.

Funding for the Sloan Digital Sky Survey IV has been provided by the Alfred P. Sloan Foundation, the U.S. Department of Energy Office of Science, and the Participating Institutions. SDSS-IV acknowledges support and resources from the Center for High-Performance Computing at the University of Utah. The SDSS web site is www.sdss.org.

SDSS-IV is managed by the Astrophysical Research Consortium for the Participating Institutions of the SDSS Collaboration including the Brazilian Participation Group, the Carnegie Institution for Science, Carnegie Mellon University, the Chilean Participation Group, the French Participation Group, Harvard-Smithsonian Center for Astrophysics, Instituto de Astrof\'isica de Canarias, The Johns Hopkins University, Kavli Institute for the Physics and Mathematics of the Universe (IPMU) / University of Tokyo, the Korean Participation Group, Lawrence Berkeley National Laboratory, Leibniz Institut f\"ur Astrophysik Potsdam (AIP), Max-Planck-Institut f\"ur Astronomie (MPIA Heidelberg), Max-Planck-Institut f\"ur Astrophysik (MPA Garching), Max-Planck-Institut f\"ur Extraterrestrische Physik (MPE), National Astronomical Observatories of China, New Mexico State University, New York University, University of Notre Dame, Observat\'ario Nacional / MCTI, The Ohio State University, Pennsylvania State University, Shanghai Astronomical Observatory, United Kingdom Participation Group, Universidad Nacional Aut\'onoma de M\'exico, University of Arizona, University of Colorado Boulder, University of Oxford, University of Portsmouth, University of Utah, University of Virginia, University of Washington, University of Wisconsin, Vanderbilt University, and Yale University.

The Legacy Surveys consist of three individual and complementary projects: the Dark Energy Camera Legacy Survey (DECaLS; NOAO Proposal ID \# 2014B-0404; PIs: David Schlegel and Arjun Dey), the Beijing-Arizona Sky Survey (BASS; NOAO Proposal ID \# 2015A-0801; PIs: Zhou Xu and Xiaohui Fan), and the Mayall z-band Legacy Survey (MzLS; NOAO Proposal ID \# 2016A-0453; PI: Arjun Dey). DECaLS, BASS and MzLS together include data obtained, respectively, at the Blanco telescope, Cerro Tololo Inter-American Observatory, National Optical Astronomy Observatory (NOAO); the Bok telescope, Steward Observatory, University of Arizona; and the Mayall telescope, Kitt Peak National Observatory, NOAO. The Legacy Surveys project is honored to be permitted to conduct astronomical research on Iolkam Du'ag (Kitt Peak), a mountain with particular significance to the Tohono O'odham Nation.

NOAO is operated by the Association of Universities for Research in Astronomy (AURA) under a cooperative agreement with the National Science Foundation.

This project used data obtained with the Dark Energy Camera (DECam), which was constructed by the Dark Energy Survey (DES) collaboration. Funding for the DES Projects has been provided by the U.S. Department of Energy, the U.S. National Science Foundation, the Ministry of Science and Education of Spain, the Science and Technology Facilities Council of the United Kingdom, the Higher Education Funding Council for England, the National Center for Supercomputing Applications at the University of Illinois at Urbana-Champaign, the Kavli Institute of Cosmological Physics at the University of Chicago, Center for Cosmology and Astro-Particle Physics at the Ohio State University, the Mitchell Institute for Fundamental Physics and Astronomy at Texas A\&M University, Financiadora de Estudos e Projetos, Fundacao Carlos Chagas Filho de Amparo, Financiadora de Estudos e Projetos, Fundacao Carlos Chagas Filho de Amparo a Pesquisa do Estado do Rio de Janeiro, Conselho Nacional de Desenvolvimento Cientifico e Tecnologico and the Ministerio da Ciencia, Tecnologia e Inovacao, the Deutsche Forschungsgemeinschaft and the Collaborating Institutions in the Dark Energy Survey. The Collaborating Institutions are Argonne National Laboratory, the University of California at Santa Cruz, the University of Cambridge, Centro de Investigaciones Energeticas, Medioambientales y Tecnologicas-Madrid, the University of Chicago, University College London, the DES-Brazil Consortium, the University of Edinburgh, the Eidgenossische Technische Hochschule (ETH) Zurich, Fermi National Accelerator Laboratory, the University of Illinois at Urbana-Champaign, the Institut de Ciencies de l'Espai (IEEC/CSIC), the Institut de Fisica d'Altes Energies, Lawrence Berkeley National Laboratory, the Ludwig-Maximilians Universitat Munchen and the associated Excellence Cluster Universe, the University of Michigan, the National Optical Astronomy Observatory, the University of Nottingham, the Ohio State University, the University of Pennsylvania, the University of Portsmouth, SLAC National Accelerator Laboratory, Stanford University, the University of Sussex, and Texas A\&M University.

BASS is a key project of the Telescope Access Program (TAP), which has been funded by the National Astronomical Observatories of China, the Chinese Academy of Sciences (the Strategic Priority Research Program ``The Emergence of Cosmological Structures'' Grant \# XDB09000000), and the Special Fund for Astronomy from the Ministry of Finance. The BASS is also supported by the External Cooperation Program of Chinese Academy of Sciences (Grant \# 114A11KYSB20160057), and Chinese National Natural Science Foundation (Grant \# 11433005).

The Legacy Survey team makes use of data products from the Near-Earth Object Wide-field Infrared Survey Explorer (NEOWISE), which is a project of the Jet Propulsion Laboratory/California Institute of Technology. NEOWISE is funded by the National Aeronautics and Space Administration.

The Legacy Surveys imaging of the DESI footprint is supported by the Director, Office of Science, Office of High Energy Physics of the U.S. Department of Energy under Contract No. DE-AC02-05CH1123, by the National Energy Research Scientific Computing Center, a DOE Office of Science User Facility under the same contract; and by the U.S. National Science Foundation, Division of Astronomical Sciences under Contract No. AST-0950945 to NOAO.

\bibliographystyle{JHEP}
\bibliography{shbk18}

\end{document}